\documentclass[prd,aps,nofootinbib,onecolumn,showpacs,10pt]{revtex4}
\usepackage{graphicx,epsfig}

\usepackage{epsf}
\usepackage{graphicx}

\def   \lsth     {littlest Higgs}

\newcommand\iden{\leavevmode\hbox{\small1\normalsize\kern-.33em1}}
\def \nn {\nonumber}
\def\w{{\rm w}}

\def\gmm{{\gamma_{\mu}}}
\def\gnn{{\gamma_{\nu}}}
\def\gff{{\gamma_{5}}}

\def\gvi{{g_{V_i}}}
\def\gai{{g_{A_i}}}
\def\gvon{{g_{V_1}}}
\def\gaon{{g_{A_1}}}

\def\mn{{g_{\mu\nu}}}

\def\mnu{{g^{\mu\nu}}}

\def\dwi{{M_i \Gamma_i}}

\def\fbg{\frac{fb}{GeV}}

\def\boss{\hskip2mm }
\def\bos{\qquad\allowbreak }

\def\bsdd{\vskip2mm}

\begin{document}
\vspace*{2cm}

\title{$Z_L$ associated pair production of charged Higgs bosons in the littlest Higgs model at $e^{+}e^{-}$ colliders  }

\author{A. \c{C}a\~{g}{\i}l\footnote{email:e110635@metu.edu.tr}, M. T. Zeyrek\footnote{email:zeyrek@metu.edu.tr}}
\affiliation{\vspace*{0.1in} Department of Physics, Middle East
Technical University,\\06531 Ankara, Turkey}

\vspace*{1.0cm}

\begin{abstract}
The production of single and doubly charged Higgs bosons associated
with standard model gauge boson $Z_L$ in $e^{+}e^{-}$ colliders are
examined. The sensitivity of these processes on the littlest Higgs
model parameters in the range of compatibility with electroweak
precision observables are analyzed. The possibility of detecting
lepton flavor violation processes are also discussed.

\end{abstract}
\pacs{12.60.-i,13.66.Fg,13.66.Hk,14.80.Cp} \maketitle



%

\section{Introduction}

One of the unsolved problems of the Standard Model(SM) is the
hierarchy problem. The little Higgs
models\cite{lh1,lhmodels1,lhmodels2,lhmodels3} are introduced to
solve the hierarchy problem by stabilizing the Higgs mass by a
collective symmetry breaking mechanism due to the cancelation of
divergent loops by appearance of new particles as a consequence of
extra symmetries. The phenomenology of the little Higgs models are
widely discussed in literature (for reviews see
\cite{perelstein1,thanrev,schmalzrev1}), and constraints on little
Higgs model parameters are studied
\cite{perelstein2ew,B1rizzo,Bdawson,Bkilian,Bdias,B2csaki}. The
little Higgs models are also expected to give new significant
signatures in future high energy colliders and studied in references
\cite{atlaswork,atlaswork2,LHCreuter,A3}, due to the new particles
which are predicted by these models. Also the $Z_L$ associated
production of SM Higgs boson at $e^+ e^-$ colliders in the framework
of the littlest Higgs model is studied in \cite{zh1,zhh2}

In this work we examined the production of single and doubly charged
scalars associated with $Z_L$ boson at future linear $e^{+}e^{-}$
colliders, namely, International Linear Collider (ILC) \cite{ILC}
and Compact Linear Collider (CLIC) \cite{CLIC} in the context of
littlest Higgs model\cite{lh1}. We examined the dependance of total
and differential cross sections of the processes to the littlest
Higgs model parameters at the range allowed by electroweak presicion
measurements. Also we analyzed the decays of scalars in the context
of lepton flavor violation. We found that the production rates of
the single charged scalar pairs associated with $Z_L$ are less than
the $Z_L$ associated production of doubly charged scalars, but both
channels will be achieved at $e^{+}e^{-}$ colliders at $\sqrt{s}\geq
2TeV$.

The layout of this paper is as follows: In section $2$ first we give
a brief review of the littlest Higgs model and then we calculate the
pair production of charged scalars at $e^{+}e^{-}$ colliders. In
this section we also calculate the total and differential cross
sections of the productions of charged scalars with $Z_L$. Section
$3$ contains our numerical results and discussions.

\section{Theoretical Framework}
Before examining the pair production of charged scalars with $Z_L$,
we introduce a few words on the littlest Higgs model based on
reference \cite{thanrev}. In the littlest Higgs model global
symmetry $SU(5)$ is broken spontaneously to $SO(5)$ at an energy
scale $f\sim 1TeV$ leaving $14$ nambu goldstone bosons(NGB)
corresponding to broken symmetries represented by the goldstone
boson matrix of non linear sigma model (nlsm) such as:
\begin{equation}\label{pi1}
   \Pi(x)=\sum_{a=1}^{14} \Pi^{a}(x) X^{a}
\end{equation}
where $X^{a}$ are broken generators of $SU(5)$. The vacuum bases
triggering the symmetry breaking can be chosen as
\begin{equation}\label{sg1}
  \Sigma_{0} =\left(
                \begin{array}{ccc}
                  0 & 0 & \iden \\
                  0 & 1 & 0 \\
                  \iden& 0 & 0 \\
                \end{array}
              \right),
\end{equation}
and the $\Sigma$ field is defined as
\begin{equation}\label{sg2}
\Sigma(x) \,=\, e^{i \Pi/f} \Sigma_0 e^{i \Pi/f} \,=\, e^{2i \Pi/f}
\Sigma_0,
\end{equation}
and the effective lagrangian of the $\Sigma$ field is:
\begin{equation}
    \mathcal{L}_{\Sigma} = \frac{f^2}{8}
    {\rm Tr} | \mathcal{D}_{\mu} \Sigma |^2.
\label{lg1}
\end{equation}
In the {\lsth} model $SU(5)$ contains the gauged subgroup
$[SU(2)_1\otimes U(1)_1]\otimes[SU(2)_2\otimes U(1)_2]$ of $SU(5)$,
defining the covariant derivative such as:
\begin{equation}\label{covder}
           \mathcal{D}_\mu \Sigma=  \partial_\mu\Sigma - i \sum_{i=1}^2\left(
g_i( W_{\mu i}\Sigma +  \Sigma W_{\mu i}^T) + g'_i (B_{\mu i}\Sigma
+ \Sigma B_{\mu i}^T) \right),
    \end{equation}
where $B_{\mu i}$ and $W_{\mu i}$ are the gauge fields, $g_{i}'$ and
$g_{i}$ are the corresponding couplings of $U(1)_{i}$ and
$SU(2)_{i}$ respectively. The vev of $\Sigma$ field in the
lagrangian breaks $(SU(2)\otimes U(1))^2$ symmetry to diagonal
subgroup $(SU(2)\otimes U(1))$ of SM.
As a consequence symmetry breaking, gauge bosons gain mass by eating
the four of the NGBs. The mixing angles between the $SU(2)$
subgroups and between the $U(1)$ subgroups are defined respectively
as:
\begin{equation}\label{ssp}
    s=\frac{g_2}{\sqrt{g_{1}^2 + g_{2}^2 }}~~,~~~~~~~~ s^\prime=\frac{g'_2}{\sqrt{g_{1}^{\prime 2} + g_{2}^{\prime
    2}
    }}~~.
\end{equation}

The usual electroweak symmetry breaking occurs by the vev of the
Higgs potential written by Coleman Weinberg method for scalars. By
EWSB vector bosons get extra mixings due to vacuum expectation
values of $h$ doublet and $\phi$ triplet. Again by diagonalizing the
mass matrices, the final masses of vector bosons to the order of
$\frac{v^2}{f^2}$ are expressed as\cite{thanrev}:
\begin{eqnarray}\label{massesvectors}
M_{W_L^{\pm}}^2 &=&  m_w^2 \left[
    1 - \frac{v^2}{f^2} \left( \frac{1}{6}
    + \frac{1}{4} (c^2-s^2)^2
    \right) + 4 \frac{v^{\prime 2}}{v^2} \right], \nn \\
        M_{W_H^{\pm}}^2 &=& \frac{f^2g^2}{4s^2c^2}
    - \frac{1}{4} g^2v^2
    + \mathcal{O}(v^4/f^2)= m_w^2\left( \frac{f^2}{s^2c^2v^2}-1\right)
    ,\nn \\
    M_{A_L}^2 &=& 0 ,\nonumber \\
    M_{Z_L}^2 &=& m_z^2
    \left[ 1 - \frac{v^2}{f^2} \left( \frac{1}{6}
    + \frac{1}{4} (c^2-s^2)^2
    + \frac{5}{4} (c^{\prime 2}-s^{\prime 2})^2 \right)
    + 8 \frac{v^{\prime 2}}{v^2} \right],
    \nonumber \\
    M_{A_H}^2 &=&
    \frac{f^2 g^{\prime 2}}{20 s^{\prime 2} c^{\prime 2}}
    - \frac{1}{4} g^{\prime 2} v^2 + g^2 v^2 \frac{x_H}{4s^2c^2}
          = m_z^2 s_{\w}^2 \left(
    \frac{ f^2 }{5 s^{\prime 2} c^{\prime 2}v^2}
    - 1 + \frac{x_H c_{\w}^2}{4s^2c^2  s_{\w}^2} \right),
    \nonumber \\
    M_{Z_H}^2 &=& \frac{f^2g^2}{4s^2c^2}
    - \frac{1}{4} g^2 v^2
    - g^{\prime 2} v^2 \frac{x_H}{4s^{\prime 2}c^{\prime 2}}
    = m_w^2 \left( \frac{f^2}{s^2c^2 v^2}
    - 1 -  \frac{x_H s_{\w}^2}{s^{\prime 2}c^{\prime 2}c_{\w}^2}\right) ,
\end{eqnarray}
where  $m_w\equiv gv/2$, $m_z\equiv {gv}/(2c_{\w})$ and $x_H =
\frac{5}{2} g g^{\prime}
    \frac{scs^{\prime}c^{\prime} (c^2s^{\prime 2} + s^2c^{\prime 2})}
    {(5g^2 s^{\prime 2} c^{\prime 2} - g^{\prime 2} s^2 c^2)}$. In these
equations $s_{\w}$ and $c_{\w}$ are the usual weak mixing angles:
\begin{eqnarray}
    s_{\w} = \frac{g^{\prime}}{\sqrt{g^2 + g^{\prime 2}}}, &\qquad&
    c_{\w} = \frac{g}{\sqrt{g^2 + g^{\prime 2}}}.
\end{eqnarray}
The parameters $v$ and $v'$ in equation \ref{massesvectors}, are the
vacuum expectation values of scalar doublet and triplet given as;
$\langle h^0 \rangle = v/\sqrt{2}$ where $v=246 GeV$ and $\langle i
\phi^0 \rangle = v^{\prime}\leq  \frac{v^2}{4f}$, bounded by
electroweak precision data. Also diagonalizing the mass matrix for
scalars the physical states are found to be the SM Higgs scalar $H$,
the neutral scalar $\phi^0$, the neutral pseudo scalar $\phi^P$, and
the charged scalars $\phi^+$ and $\phi^{++}$. The masses of these
scalars are degenerate, and in terms of Higgs mass can be expressed
as:
\begin{eqnarray}
    M_\phi \approx\frac{\sqrt{2}M_H f}{v}.
\end{eqnarray}
The scalar fermion interactions in the model are written in yukawa
lagrangian preserving gauge symmetries of the model for SM leptons
and quarks, including the third generation having an extra singlet,
the $T$ quark. In this work, leptons are charged under both $U(1)$
groups, with corresponding hypercharges of $Y_1$ and $Y_2$. The
restriction for $Y_1$ and $Y_2$ is that $Y_1 + Y_2$ should reproduce
$U(1)_Y$ hypercharge $Y$ of SM, thus $Y_1=x Y$ and $Y_2=(1-x)Y$ can
be written. Due to gauge invariance, $x$ can be taken as
$3/5$\cite{thanrev,B2csaki}. Also for light fermions, a Majorano
type mass term can be implemented in yukawa
lagrangian\cite{thanlept1,gaurlept1,cinlept_L2yue,gaurlept2} which
results in lepton flavor violation by unit two, such as:
\begin{equation}\label{lepviol1}
    {\cal L}_{LFV} = iY_{ij} L_i^T \ \phi \, C^{-1} L_j + {\rm h.c.},
\end{equation}
where $L_i$ are the lepton doublets $\left(
                                       \begin{array}{cc}
                                         l &\nu_l \\
                                       \end{array}
                                     \right)$,
and $Y_{ij}$ is the yukawa coupling with $Y_{ii}=Y$ and $Y_{ij(i\neq
j)}=Y'$ . The values of yukawa couplings $Y$ and $Y'$ are restricted
by the current constraints on the neutrino
masses\cite{neutrinomass}, given as; $M_{ij}=Y_{ij}v'\simeq
10^{-10}GeV$\cite{thanlept1}. Since the vacuum expectation value
$v'$ has only an upper bound; $v'<1GeV$, $Y_{ij}$ can be taken up to
order of unity without making $v'$ unnaturally small.

In the {\lsth} model the symmetry breaking scale $f$, and the mixing
angles $s,s'$ are not restricted by the model. These parameters are
constrained by observables of electroweak presicion
data\cite{perelstein2ew,B1rizzo,Bdawson,Bkilian,Bdias,B2csaki}. For
the values of the symmetry breaking scale $1TeV \leq f \leq 2 TeV$,
mixing angles are in the range $0.75\leq s \leq 0.99$ and $0.6\leq
s' \leq 0.75 $, for $ 2TeV \leq f \leq 3TeV$ they have acceptable
values in the range $0.6\leq s \leq 0.99$ and $0.6\leq s' \leq 0.8$,
for $3TeV \leq f \leq 4 TeV$ they are in the range  $0.4\leq s \leq
0.99$ and $0.6\leq s' \leq 0.85$, and for the higher values of the
symmetry breaking scale, i.e. $f\geq 4TeV$, the mixing angles are
less restricted and they are in the range  $0.15\leq s \leq 0.99$
and $0.4\leq s' \leq 0.9$ \cite{B2csaki}.

After these preliminary remarks lets start to calculate the cross
sections of the pair productions for the charged scalars with $Z_L$
at $e^{+}e^{-}$ colliders. In the {\lsth} model there are four
neutral vector bosons; the SM $Z$ boson $Z_ L$, massless photon
$A_L$, and two new bosons $Z_H$ and $A_H$. Their couplings to
fermions are written as $i \gmm (\gvi + \gai \gff)$ where
$i=1,2,3,4$ corresponds to $Z_L$, $Z_H$, $A_H$ and $A_L$
respectively. The couplings of gauge vector to electron positron
pairs are given in table \ref{gVgA}, where $e=\sqrt{4 \pi\alpha}$,
$y_e=\frac{3}{5}$ for anomaly cancelations, $x_Z^{W^{\prime}} =
-\frac{1}{2c_{\w}} sc(c^2-s^2)$ and $x_Z^{B^{\prime}} =
-\frac{5}{2s_{\w}} s^{\prime}c^{\prime}
    (c^{\prime 2}-s^{\prime 2})$. It is seen that vector and axial vector couplings
of SM $Z_L e^{+}e^{-}$ vertex also gets contributions from {\lsth}
model. As a result total decay widths of SM vector bosons also gets
corrections of the order $\frac{v^2}{f^2}$, since the decay widths
of vectors to fermion couples are written as; $\Gamma(V_i\rightarrow
f\bar{f})=\frac{N}{24\pi}(g^2_V+g^2_A)M_{V_i}$ where $N=3$ for
quarks, and $N=1$ for fermions. The total decay widths of the new
vectors are given as \cite{A3}:
\begin{eqnarray}
\nn \Gamma_{A_H}&\approx & \frac{g'^2 M_{A_H}(21-70 s'^2 +59
s'^4)}{48 \pi s'^2 (1-s'^2) },\\
\nn\Gamma_{Z_{H}}&\approx& \frac{g^2 (193 - 388 s^2 + 196 s^4)}{768
\pi s^2 (1-s^2)}M_{Z_H},\\
\Gamma_{W_H}&\approx& \frac{g^2(97-196 s^2+100 s^4)}{384\pi s^2
(1-s^2 )}M_{W_H}.
\end{eqnarray}
The new scalars and pseudo scalars also contribute to the analysis
done in this study. Since these new scalars have lepton flavor
violating decay modes, their total widths will depend on the yukawa
couplings $Y_{ii}=Y$ and $Y_{ij(i\neq j)}=Y'$ if the flavor
violating term in the yukawa lagrangian is considered. The decay
widths of scalars are given as\cite{thanlept1}:
\begin{equation}\label{dwp2}
 \nn   \Gamma_{\phi}\approx  \frac{v^{\prime 2} M_{\phi}^3}{2 \pi v^4}+\frac{3}{8\pi }
|Y|^2 M_\phi+\frac{3}{4\pi } |Y'|^2 M_\phi ,
\end{equation}
where we neglected the decays into quarks since they are very small
compared to other terms. The first term in the decay width
proportional to $|Y|^2$ corresponds to decays into same family of
leptons and the second term proportional to $|Y'|^2$ correspond to
the decays into different families. The branching ratios of scalars
decaying to same family of leptons is denoted as $BR[Y]$ and to
leptons of different flavor is $BR[Y']$. In this work the values of
the yukawa mixings are taken to be $10^{-4}\leq Y\leq 1$ and $Y'\leq
10^{-4}$.
%
%
\begin{table}[hbt]
\begin{center}
\begin{tabular}{|c||c|c|c|}
  \hline
   $i$& vertices &$\gvi$  &$\gai$    \\
  \hline\hline 1&$e\bar{e} Z_L$ &
    $-\frac{g}{2c_{\w}} \left\{ (-\frac{1}{2} + 2 s^2_{\w})- \frac{v^2}{f^2} \left[ -c_{\w} x_Z^{W^{\prime}} c/2s
    \right. \right.$ &
    $-\frac{g}{2c_{\w}} \left\{ \frac{1}{2}
    - \frac{v^2}{f^2} \left[ c_{\w} x_Z^{W^{\prime}} c/2s
    \right. \right.$ \\
& &  $\left. \left.
    + \frac{s_{\w} x_Z^{B^{\prime}}}{s^{\prime}c^{\prime}}
    \left( 2y_e - \frac{9}{5} + \frac{3}{2} c^{\prime 2}
    \right) \right] \right\}$&
    $\left. \left.
    + \frac{s_{\w} x_Z^{B^{\prime}}}{s^{\prime}c^{\prime}}
    \left( -\frac{1}{5} + \frac{1}{2} c^{\prime 2} \right)
    \right] \right\}$ \\
  \hline 2& $e\bar{e} Z_H$ &$-gc/4s$ & $gc/4s$      \\
  \hline 3&$e\bar{e} A_H$  &
    $\frac{g^{\prime}}{2s^{\prime}c^{\prime}}
    \left( 2y_e - \frac{9}{5} + \frac{3}{2} c^{\prime 2} \right)$ &
    $\frac{g^{\prime}}{2s^{\prime}c^{\prime}}
    \left( -\frac{1}{5} + \frac{1}{2} c^{\prime 2} \right)$    \\
  \hline4& $e\bar{e} A_L$&$e$&0     \\
  \hline

\end{tabular}
\caption{The vector and axial vector couplings of $e\bar{e}$ with
vector bosons. Feynman rules for $e\bar{e} V_i$ vertices are given
as $i \gmm (\gvi + \gai \gff)$\cite{thanrev} .}\label{gVgA}
\end{center}

\end{table}
\begin{table}[htb]
\begin{center}
\begin{tabular}{|c||c|c|}
  \hline
  i/j&vertices & $i C^{\phi\phi}_{ij} \mn$ \\
\hline\hline  1/1 &$\phi^+ \phi^- Z_L Z_L$& $ 2 i \frac{g^2}{c^2_\w} s^4_\w g_{\mu\nu}$ \\
\hline  2/1&$\phi^+ \phi^- Z_H Z_L$ & $\mathcal{O}(v^2/f^2)\sim 0$ \\
\hline  3/1&$\phi^+ \phi^- A_H Z_L$ &$2 i \frac{gg^{\prime}}{c_\w}
    \frac{(c^{\prime 2}-s^{\prime 2})}{2s^{\prime}c^{\prime}}
        s^2_\w g_{\mu\nu}$  \\
\hline  4/1&$\phi^+ \phi^- A_L Z_L$ &$- 2 i e \frac{g}{c_\w} s^2_\w g_{\mu\nu}$   \\
\hline
\end{tabular}

\caption{Feynman rules for the four point  $\phi^+ \phi^- V_i V_j$
vertices\cite{thanrev}.} \label{PP11VVcouplings}
\end{center}
\end{table}
\begin{table}[htb]
\begin{center}
\begin{tabular}{|c||c|c|}
  \hline
  i/j&vertices & i $E^{\phi\phi}_{i} Q_\mu$\\
\hline\hline  1 &$\phi^+ \phi^- Z_L$&$i \frac{g}{c_\w} s^2_\w (p_1-p_2)_{\mu}$   \\
\hline  2&$\phi^+ \phi^- Z_H$ &$\mathcal{O}(v^2/f^2)\sim 0$  \\
\hline  3&$\phi^+ \phi^- A_H$ & $i
    g^{\prime} \frac{(c^{\prime 2}-s^{\prime 2})}{2s^{\prime}c^{\prime}}
    (p_1-p_2)_{\mu}$ \\
\hline  4&$\phi^+ \phi^- A_L$ & $-i e (p_1 - p_2)_{\mu}$ \\
\hline
\end{tabular}

\caption{Feynman rules for $\phi^+ \phi^- V_i $
vertices\cite{thanrev}.} \label{PP11Vcouplings}
\end{center}
\end{table}
\begin{table}[htb]
\begin{center}
\begin{tabular}{|c||c|c|}
  \hline
  i/j&vertices & $ i B^{\phi W}_{ij} \mn$ \\
\hline\hline  1/1 &$\phi^+ W^-_L Z_L$& $-i \frac{g^2}{c_\w} v^{\prime} g_{\mu\nu}$\\
\hline 1/2 & $\phi^+ W^-_H Z_L$ & $ i \frac{g^2}{c_\w} \frac{(c^2-s^2)}{2sc} v^{\prime} g_{\mu\nu}$ \\
\hline  2/1&$\phi^+ W^-_L Z_H$ & $ i g^2 \frac{(c^2-s^2)}{2sc} v^{\prime} g_{\mu\nu}$\\
\hline 2/2&$\phi^+ W^-_H Z_H$ & $-i g^2 \frac{(c^4+s^4)}{2s^2c^2} v^{\prime} g_{\mu\nu}$\\
\hline  3/1&$\phi^+ W^-_L A_H$ & $-\frac{i}{2} g g^{\prime}
    \frac{(c^{\prime 2}-s^{\prime 2})}{2s^{\prime}c^{\prime}}
    ( v s_+ - 4 v^{\prime} ) g_{\mu\nu}$\\
\hline 3/2&$\phi^+ W^-_H A_H$ & $-\frac{i}{2} g g^{\prime}
    \frac{(c^2c^{\prime 2}+s^2s^{\prime 2})}{scs^{\prime}c^{\prime}}
    v^{\prime} g_{\mu\nu}$\\
\hline
\end{tabular}

\caption{The Feynman rules for $\phi^+ V_i W^-_j$ vertices. Their
couplings are given in the form $i B^{\phi W}_{ij} \mn$ where $\mn$
carries the lorentz indices of vectors and $j=1,2$ denotes $W_L ,
W_H$ respectively\cite{thanrev}.} \label{P1WVcouplings}
\end{center}
\end{table}
\begin{table}[htb]
\begin{center}
\begin{tabular}{|c||c|c|}
  \hline
  i/j&vertices & $i C'^{\phi\phi}_{ij}\mn$ \\
\hline\hline  1/1 &$\phi^{++} \phi^{--} Z_L Z_L$& $2 i \frac{g^2}{c^2_\w} (1 - 2 s^2_\w)^2 g_{\mu\nu}$  \\
\hline  2/1&$\phi^{++} \phi^{--} Z_H Z_L$ &$2 i \frac{g^2}{c_\w}
\frac{(c^2-s^2)}{2sc}
        ( 1 - 2 s^2_\w ) g_{\mu\nu}$  \\
\hline  3/1&$\phi^{++} \phi^{--} A_H Z_L$ &$-2 i
\frac{gg^{\prime}}{c_\w}
    \frac{(c^{\prime 2}-s^{\prime 2})}{2s^{\prime}c^{\prime}}
        ( 1 - 2 s^2_\w ) g_{\mu\nu}$  \\
\hline  4/1&$\phi^{++} \phi^{--} A_L Z_L$ & $4 i e \frac{g}{c_\w} (1 - 2 s^2_\w) g_{\mu\nu}$ \\
\hline
\end{tabular}

\caption{Feynman rules for the four point  $\phi^{++} \phi^{--} V_i
V_j$ vertices\cite{thanrev}.} \label{PP22VVcouplings}
\end{center}
\end{table}
\begin{table}[htb]
\begin{center}
\begin{tabular}{|c||c|c||c||c|c|}
  \hline
  i/j&vertices & $i E'^{\phi\phi}_{i} Q_\mu$ \\
\hline\hline  1 &$\phi^{++} \phi^{--}Z_L$&$-i \frac{g}{c_\w} \left( 1 - 2 s^2_\w \right) (p_1-p_2)_{\mu}$   \\
\hline  2&$\phi^{++} \phi^{--}Z_H$ &$i g \frac{(c^2-s^2)}{2sc} (p_1-p_2)_{\mu}$  \\
\hline  3&$\phi^{++} \phi^{--} A_H$ &$i
    g^{\prime} \frac{(c^{\prime 2}-s^{\prime 2})}{2s^{\prime}c^{\prime}}
    (p_1-p_2)_{\mu}$  \\
\hline  4&$\phi^{++} \phi^{--} A_L$ &$-2 i e (p_1-p_2)_{\mu}$  \\
\hline
\end{tabular}

\caption{Feynman rules for $\phi^{++} \phi^{--} V_i $
vertices\cite{thanrev}.} \label{PP22Vcouplings}
\end{center}
\end{table}

\begin{figure}[tbh]
\begin{center}
\includegraphics[width=3.5cm]{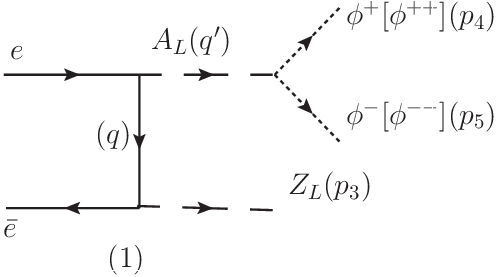}\boss\includegraphics[width=3.5cm]{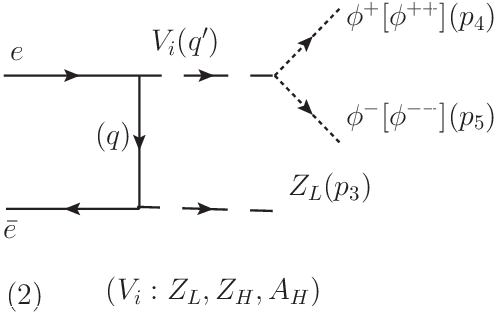}\boss\includegraphics[width=3.5cm]{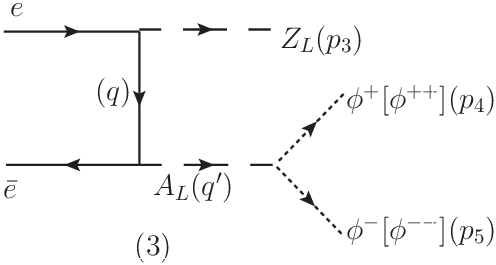}
\bsdd
\includegraphics[width=3.5cm]{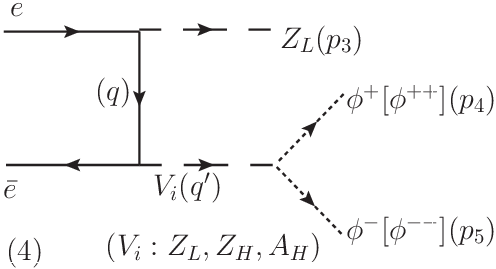}\boss\includegraphics[width=3.5cm]{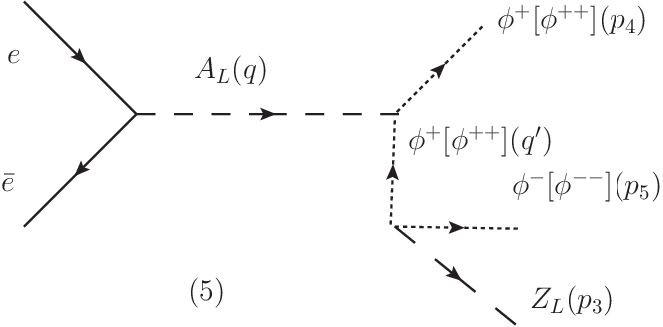}\boss\includegraphics[width=3.5cm]{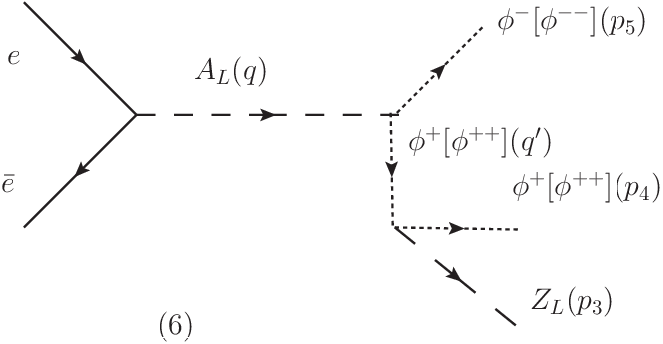}
\bsdd
\includegraphics[width=3.5cm]{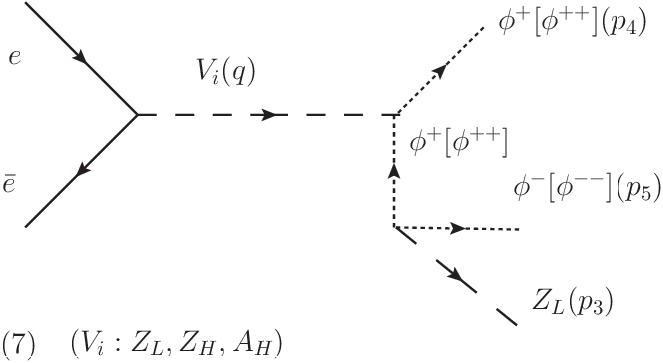}\boss\includegraphics[width=3.5cm]{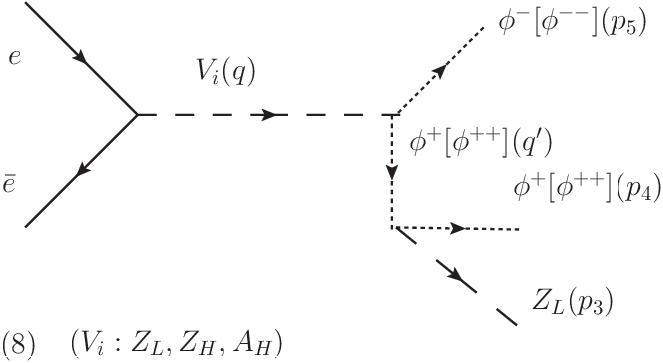}\boss\includegraphics[width=3.5cm]{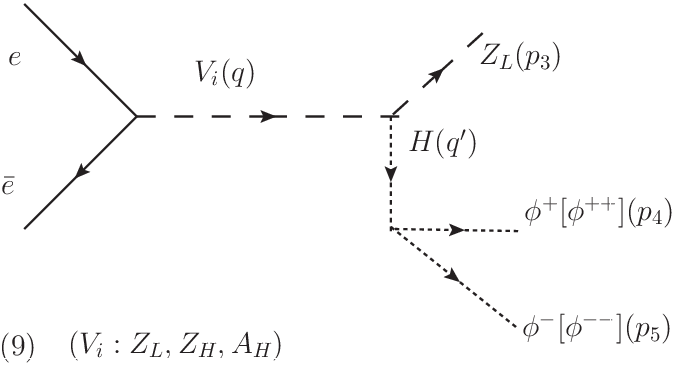}
\bsdd
\includegraphics[width=3.5cm]{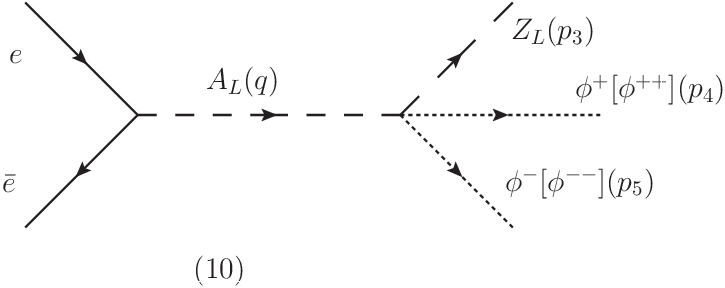}\boss\includegraphics[width=3.5cm]{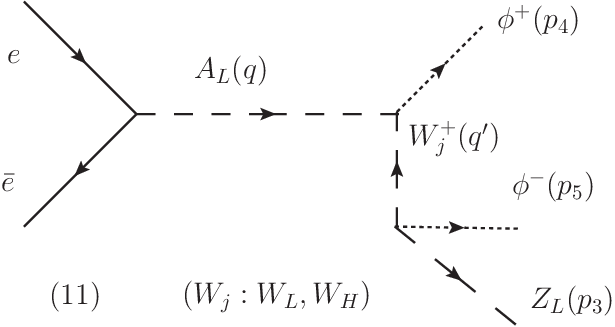}\boss\includegraphics[width=3.5cm]{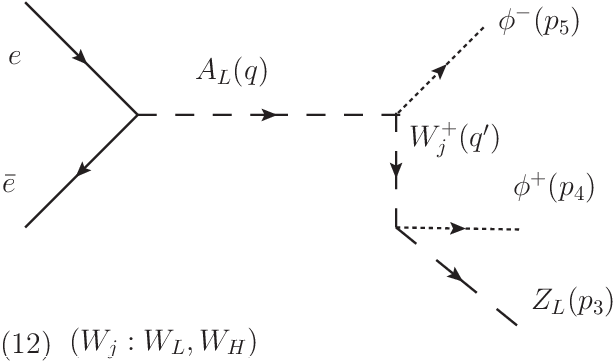}
\bsdd
\boss\includegraphics[width=3.5cm]{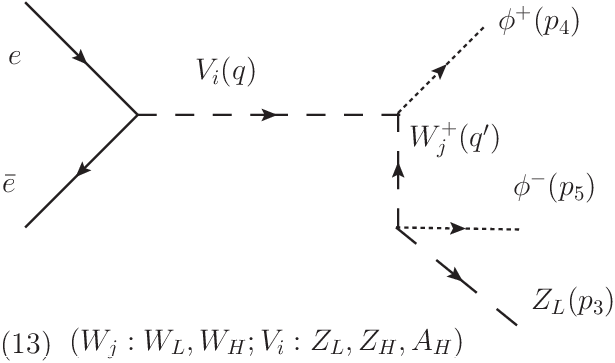}\boss\includegraphics[width=3.5cm]{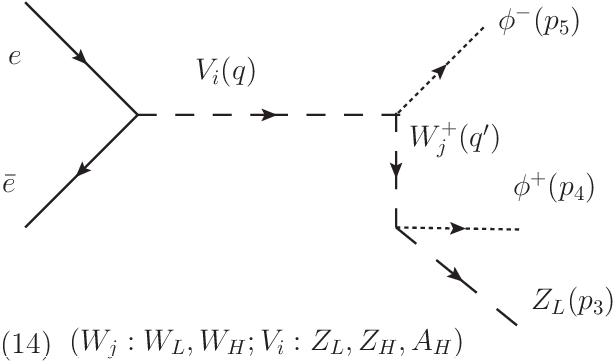}

\qquad\allowbreak \vskip2mm
\end{center}\caption{Feynman diagrams
contributing to $e^{+}e^{-}\rightarrow Z_L \phi^+ \phi^-$ and
$e^{+}e^{-}\rightarrow Z_L \phi^{++} \phi^{--}$ in {\lsth} model.
The last four diagrams only contribute to the single charged
scalars.} \label{fdx}
\end{figure}

The Feynman diagrams for the $Z_L$ associated scalar pair production
processes are given in figure \ref{fdx}. The first ten of the
Feynman diagrams in figure \ref{fdx} are for the both processes
$e^{+}e^{-}\rightarrow Z_L \phi^+ \phi^- $ and
$e^{+}e^{-}\rightarrow Z_L \phi^{++} \phi^{--} $. The last four ones
only contribute to the associated production of single charged
scalars. Now we present the amplitudes for these Feynman diagrams.
The relevant Feynman rules for the vertices for single charged
scalars are given in tables \ref{PP11VVcouplings},
\ref{PP11Vcouplings} and \ref{P1WVcouplings} and for the doubly
charged scalars in tables \ref{PP22VVcouplings} and
\ref{PP22Vcouplings}. The amplitudes for the associated production
of doubly charged scalars are get by replacing the couplings with
the primed ones. In the calculations of cross sections, terms are
expanded up to order of $v^2/f^2$.

The first four Feynman diagrams contributing to both processes
contain only the $e^{+}e^{-}V_i$ vertices. The amplitudes for the
first two diagrams in figure \ref{fdx}, are written as:
\begin{equation}\label{Mc1i}
  M_{1}=\bar{u}[-p_2] i \gmm (\gvon + \gaon \gff)\epsilon^{\mu}[p_3]\frac{i\not{q}}{q^2}i \gnn g_{V_4}u[p_1](i)
    \frac{g^{\alpha\nu}}{q'^2} i E^{\phi\phi}_{4} (p_4 - p_5)_{\alpha},
\end{equation}
\begin{eqnarray}\label{Mc2i}
\nn  M_{2}&=&\sum_{i=1}^{3}\bar{u}[-p_2] i \gmm (\gvon + \gaon
\gff)\epsilon^{\mu}[p_3]\frac{i\not{q}}{q^2}i \gnn (\gvi +\gai
\gff)u[p_1] (i)\\
& &
\frac{g^{\alpha\nu}-\frac{q'^{\nu}q'^{\alpha}}{M^{2}_{i}}}{q'^2-M^{2}_{i}+i\dwi}
i E^{\phi\phi}_{i} (p_4 - p_5)_{\alpha},
\end{eqnarray}
where $q=p_2-p_3$, $q'=p_4+p_5$ and coefficients $E^{\phi\phi}_{i}$
are given in table \ref{PP11Vcouplings}($E'^{\phi\phi}_{i}$ for
doubly charged scalars given in table \ref{PP22Vcouplings}). The
amplitudes for diagrams $3$ and $4$ in figure \ref{fdx}, are written
as:
\begin{equation}\label{Mc3i}
    M_{3}=\bar{u}[-p_2] i \gmm (g_{V_4})\frac{i \mnu}{q^2}i E^{\phi\phi}_{4}(p_4 - p_5)_{\nu} \gamma_{\alpha}
    (\gvon +\gaon \gff)\epsilon^{\alpha}[p_3]u[p_1],
\end{equation}
\begin{eqnarray}\label{Mc4i}
 \nn M_{4}&=&\sum_{i=1}^{3}\bar{u}[-p_2] i \gmm (\gvi +\gai \gff)i \frac{g^{\mu\nu}-\frac{q'^{\mu}q'^{\nu}}{M_{i}^2}}{q'^2 -
   M_{i}^2+i\dwi
    }i E^{\phi\phi}_{i}(p_4 - p_5)_{\nu}\\
    & &i\frac{\not{q}}{q^2}i\gamma_{\alpha}(\gvon + \gaon \gff)
\epsilon^{\alpha}[p_3]u[p_1],
\end{eqnarray}
where $q=p_1-p_3$, $q'=p_4+p_5$ and coefficients
$E^{\phi\phi}_{i}$($E'^{\phi\phi}_{i}$) are given in table
\ref{PP11Vcouplings}(\ref{PP22Vcouplings}). The rest of the
amplitudes from $5$ to $14$ in figure \ref{fdx} are $s$ channel
processes with $q=p_1+p_2$. For the diagrams $5$ to $8$ in figure
\ref{fdx}, corresponding amplitudes $5$ to $8$ have a heavy scalar
as a propagator, and proportional to the square of the couplings of
two vectors with a charged scalar. These amplitudes are written as:
\begin{equation}\label{Mc9}
 M_{5}=\bar{u}[-p_2] i \gmm g_{V_4}u[p_1]i \frac{\mnu}{q^2}iE^{\phi\phi}_4 (p_4 - q')_{\nu}
     \frac{i}{q'^2-M_{\phi^2}}iE^{\phi\phi}_1 (-q'-p_3)_{\sigma}\epsilon^{\sigma}[p_3],
\end{equation}
\begin{equation}\label{Mc10}
 M_{6}=\bar{u}[-p_2] i \gmm g_{V_4}u[p_1]i \frac{ \mnu}{q^2}iE^{\phi\phi}_4 (p_5 + q')_{\nu}
     \frac{i}{q'^2-M_{\phi^2}}iE^{\phi\phi}_1 (-q'-p_3)_{\sigma}\epsilon^{\sigma}[p_3],
\end{equation}
\begin{eqnarray}\label{Mc11j}
 \nn M_{7}&=&\sum_{i=1}^{3}\bar{u}[-p_2] i \gmm (\gvi +\gai \gff)u[p_1]i \frac{g^{\mu\nu}-\frac{q^{\mu}q^{\nu}}{M_{i}^2}}{q^2 -
  M_{i}^2+i\dwi
    } (i E^{\phi\phi}_i)(p_4-q')_{\nu}\\
    & & \frac{i}{q'^2 - M_{\phi}^2}
    (i E^{\phi\phi}_1)(-q'-p_3)_{\sigma}\epsilon^{\sigma}[p_3],
\end{eqnarray}
\begin{eqnarray}\label{Mc12j}
  \nn M_{8}&=&\sum_{i=1}^{3}\bar{u}[-p_2] i \gmm (\gvi +\gai \gff)u[p_1]i \frac{g^{\mu\nu}-\frac{q^{\mu}q^{\nu}}{M_{i}^2}}{q^2 -
  M_{i}^2+i\dwi
    } (i E^{\phi\phi}_i)(p_5+q')_{\nu}\\
    & & \frac{i}{q'^2 - M_{\phi}^2}
    (i E^{\phi\phi}_1)(-q'-p_3)_{\sigma}\epsilon^{\sigma}[p_3],
\end{eqnarray}
where $E^{\phi\phi}_i$($E'^{\phi\phi}_i$) are given in table
\ref{PP11Vcouplings}(\ref{PP22Vcouplings}), and $q'=q-p_5$ for
amplitudes $6$ and $8$ and $q'=q-p_4$ for $5$ and $7$.
The diagrams $9$ and $10$ in figure \ref{fdx} are sub processes in
which the four point couplings of vectors and scalars contribute.
Their amplitudes are given as:
\begin{equation}\label{Mc14}
    M_{9}=\bar{u}[-p_2] i \gmm g_{V_4}u[p_1]i \frac{i \mnu}{q^2}iC^{\phi\phi}_{41} g_{\nu\sigma}\epsilon^{\sigma}[p_3],
\end{equation}
\begin{equation}\label{Mc15j}
    M_{10}=\sum_{i=1}^{3}\bar{u}[-p_2] i \gmm (\gvi +\gai \gff)u[p_1]i \frac{g^{\mu\nu}-\frac{q^{\mu}q^{\nu}}{M_{i}^2}}{q^2 -
    M_{i}^2+i\dwi
    } (i C^{\phi\phi}_{i1})g_{\nu\alpha}\epsilon^{\alpha}[p_3],
\end{equation}
where $C^{\phi\phi}_{ij}$($C'^{\phi\phi}_{ij}$) are four point
couplings given in table
\ref{PP11VVcouplings}(\ref{PP22VVcouplings}).
The amplitudes $11$ to $14$ corresponds to diagrams $11$ to $14$ in
figure \ref{fdx}  are only for single charged scalars, where an
electric charge is carried by a $W$ propagator. These sub processes
are proportional to the couplings of a vector with a charged scalar
and a charged vector $W_j$, where $j=1,2$ corresponds to $W_L$ and
$W_H$ respectively. Their amplitudes are given as:
\begin{equation}\label{Mc5i}
 M_{11}=\sum_{j=1}^{2}\bar{u}[-p_2] i \gmm (g_{V_4})u[p_1]\frac{i \mnu}{q^2}(i B^{\phi W}_{4j}) g_{\nu\alpha}i
    \frac{g^{\alpha\beta}-\frac{q'^{\alpha}q'^{\beta}}{M_{W_j}^2}}{q^2 - M_{W_j}^2+iM_{W_j}\Gamma_{W_j}}
     (i B^{\phi W}_{1j})g_{\beta\sigma}\epsilon^{\sigma}[p_3],
\end{equation}
\begin{equation}\label{Mc6i}
   M_{12}=\sum_{j=1}^{2}\bar{u}[-p_2] i \gmm (g_{V_4})u[p_1]\frac{i \mnu}{q^2}(i B^{\phi W}_{4j}) g_{\nu\alpha}i
    \frac{g^{\alpha\beta}-\frac{q'^{\alpha}q'^{\beta}}{M_{W_j}^2}}{q^2 - M_{W_j}^2+iM_{W_j}\Gamma_{W_j}}
    (i B^{\phi W}_{1j})g_{\beta\sigma}\epsilon^{\sigma}[p_3],
\end{equation}
\begin{eqnarray}\label{Mc7ij}
\nn M_{13}&=&\sum_{i,j=1,1}^{3,2}\bar{u}[-p_2] i \gmm (\gvi +\gai
\gff)u[p_1]i \frac{g^{\mu\nu}-\frac{q^{\mu}q^{\nu}}{M_{i}^2}}{q^2 -
  M_{i}^2+i\dwi
    }(i B^{\phi W}_{ij})\\
    & & g_{\nu\alpha}
     \frac{g^{\alpha\beta}-\frac{q'^{\alpha}q'^{\beta}}{M_{W_j}^2}}{q^2 - M_{W_j}^2+iM_{W_j}\Gamma_{W_j}}
    (i B^{\phi W}_{1j})g_{\beta\sigma}\epsilon^{\sigma}[p_3],
\end{eqnarray}
\begin{eqnarray}\label{Mc8ij}
\nn M_{14}&=&\sum_{i,j=1,1}^{3,2}\bar{u}[-p_2] i \gmm (\gvi +\gai
\gff)u[p_1]i \frac{g^{\mu\nu}-\frac{q^{\mu}q^{\nu}}{M_{i}^2}}{q^2 -
    M_{i}^2+i\dwi
    }(i B^{\phi W}_{ij})\\
    & & g_{\nu\alpha}\frac{g^{\alpha\beta}-\frac{q'^{\alpha}q'^{\beta}}{M_{W_j}^2}}{q^2 - M_{W_j}^2+iM_{W_j}\Gamma_{W_j}}
     (i B^{\phi W}_{1j})g_{\beta\sigma}\epsilon^{\sigma}[p_3],
\end{eqnarray}
where $B^{\phi W}_{ij}$ are given in table \ref{P1WVcouplings}, and
$q'=q-p_5$ for amplitudes $11$ and $13$ and $q'=q-p_4$ for $12$ and
$14$.

The numerical calculations of cross sections of the pair production
of charged scalars are performed by {\tt CalcHep} generator
\cite{calchep}.
%
\section{Results and discussions}
%
%
In this section we present and discuss our results for the $Z_L$
associated pair production of charged scalars. In performing the
numerical calculations, we take the electromagnetic coupling
constant $e=\sqrt{4\pi \alpha}=0.092$, the Higgs mass $M_H=120GeV$
and the mass of the standard model bosons $M_{Z_L}=91GeV$,
$M_{W_L}=80GeV$ and the SM mixing angle $s_W=0.47$ using the recent
data\cite{pdg}. In the calculations, we ignored $v^2/f^2$ terms in
the couplings, since we are not dealing with the corrections to a SM
process.

\begin{figure}[tbh]
\begin{center}
\includegraphics[width=6.5cm]{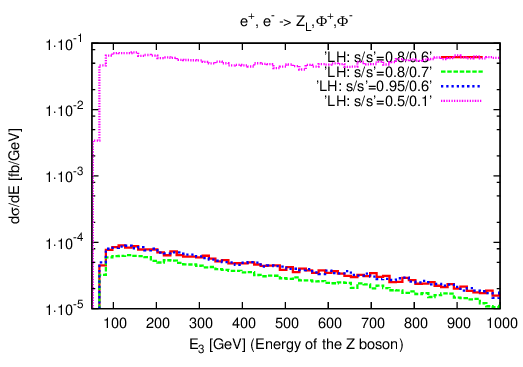}\bos\includegraphics[width=6.5cm]{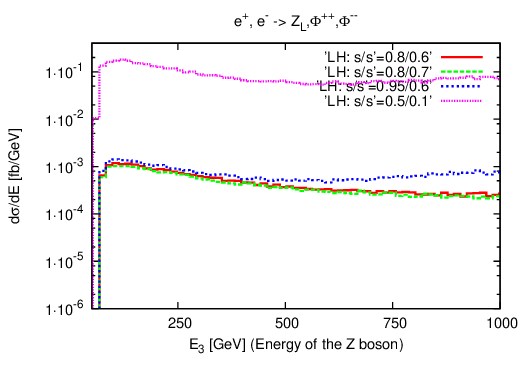}
\vskip2mm ~~\hskip10mm (a)\hskip60mm (b) \vskip2mm
\caption{Differential cross section vs. $E_Z$ graphs for processes
$e^{+}e^{-}\rightarrow Z_L \phi^+ \phi^-$(a) and
$e^{+}e^{-}\rightarrow Z_L \phi^{++} \phi^{--}$(b) for the fixed
value of parameters $s/s':0.8/0.6,0.8/0.7,0.95/0.6,0.5/0.1$ at
$f=1000GeV$ at $\sqrt{s}=3TeV$.} \label{zppC1a}
\end{center}
\end{figure}

\begin{figure}[htb]
\begin{center}
\includegraphics[width=6.5cm]{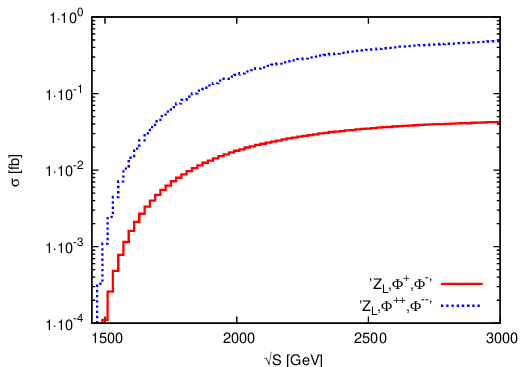}
\vskip2mm \caption{Total cross section vs. $\sqrt{s}$ graphs for
$Z_L$ associated pair production of charged scalars at $f=1TeV$ when
parameters $s/s':0.8/0.7$.} \label{PCM12}
\end{center}
\end{figure}

For the pair production of single charged heavy scalars associated
with $Z_L$ , the differential cross sections versus energy of the
$Z_L$ boson graphs for fixed values of mixing angle parameters
$s/s'$ at symmetry breaking scale $f=1TeV$ at total center of mass
energy $\sqrt{s}=3TeV$ appropriate for CLIC are presented in figure
\ref{zppC1a}(a). The total cross sections for these processes for
different fixed values of the parameters
$s/s'=0.8/0.6,0.8/0.7,0.95/0.6,0.5/0.1$ are presented in table
\ref{cseccharged}. The dependance of the total cross section on
$\sqrt{s}$ for these processes are presented in figure \ref{PCM12}.
The differential cross section gets its maximum value about
$0.1\fbg$ when $s/s'=0.5/0.1$, correspondingly for total cross
section we obtain that $59fb$. This means that thousands of
productions per year at high integrated luminosity of $100 fb^{-1}$
are expected. But this parameter set is not allowed when symmetry
breaking scale $f=1TeV$ by electroweak precision data.
\begin{table}[htb]
\begin{center}
\begin{tabular}{|c||c|c|}
  \hline $s/s'$  &$\sigma_{Z_L\phi^+\phi^-}$  & $\sigma_{Z_L\phi^{++}\phi^{--}}$ \\
  \hline
  \hline  $0.8/0.6$&$0.042$  & $ 0.48$ \\
  \hline  $0.8/0.7$& $0.031 $&  $0.44$ \\
  \hline  $0.95/0.6$&$0.043$  &$ 0.78$  \\
  \hline  $0.5/0.1$& $59$ &  $84$ \\
  \hline
\end{tabular}
\caption{The total cross sections in $fb$ for pair production of
charged scalars associated with $Z_L$ for $f=1TeV$ and at
$\sqrt{s}=3TeV$.} \label{cseccharged}
\end{center}
\end{table}
For parameters $s/s'=0.8/0.6,0.8/0.7,0.95/0.6$, the peak values of
differential cross sections are obtained at the order of
$10^{-4}\fbg$ for low $E_Z$ values, $E_Z\sim 100GeV$. The total
cross section is calculated as $0.04 fb$. This result implies
$1\sim10$ events per year accessible for a collider luminosity of
$100fb^{-1}$. The single charged scalar $\phi^-$ has leptonic decay
modes $l_i \nu_i$ and $l_i \nu_j$ with branching ratios proportional
to $|Y^2|$ and $|Y'^2|$ respectively. It also has decay modes to SM
particles; $W^\pm_L H$, $W^\pm_L Z_L$ pairs. The SM decays dominates
when the yukawa coupling $Y$ is either zero or very small ($Y\ll
1$). For higher values of yukawa coupling ($Y\sim 1$) the leptonic
modes ($\l_i \nu_i$) of $\phi^+\phi^-$ couple dominates the final
states, giving a signature of missing energy of the neutrino anti
neutrino couple, and a lepton anti lepton pair of same or different
flavor plus $Z_L$. These signals can be lepton flavor violating but
the observation is challenging due to the low production rates of
the single charged pair.

\begin{figure}[tbh]
\begin{center}
\includegraphics[width=6.5cm]{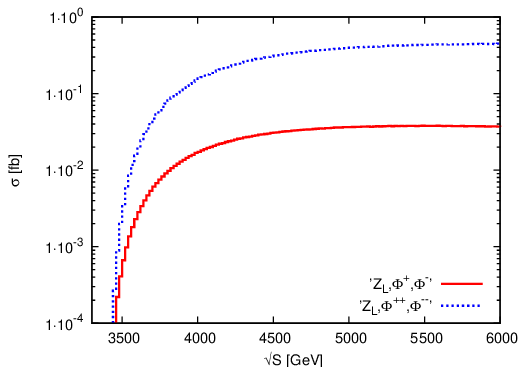}\bos\includegraphics[width=6.5cm]{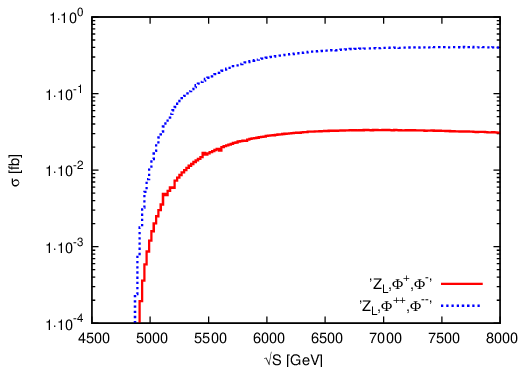}
\vskip2mm ~~\hskip10mm (a)\hskip60mm (b) \vskip2mm \caption{Total
cross section vs. $\sqrt{s}$ graphs for $Z_L$ associated pair
production of charged scalars at (a)$f=2.5TeV$, and at (b)$f=3.5TeV$
when parameters $s/s':0.8/0.7$.} \label{PCM12x2}
\end{center}
\end{figure}
\begin{figure}[tbh]
\begin{center}
\includegraphics[width=6.5cm]{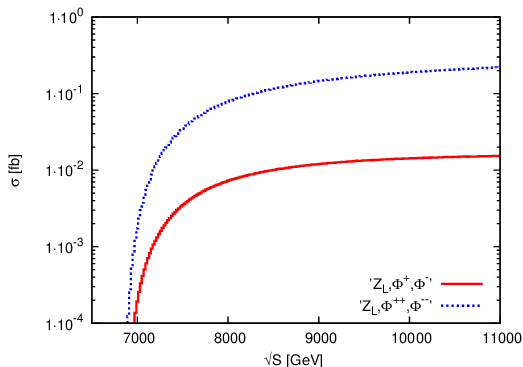}\bos\includegraphics[width=6.5cm]{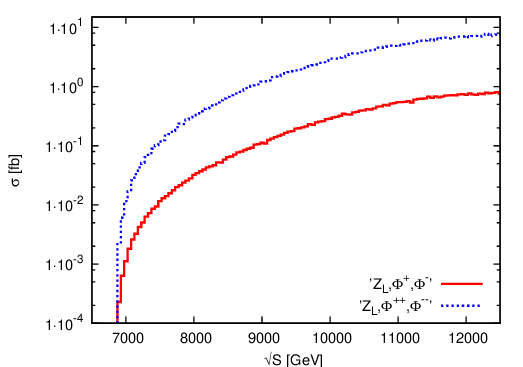}
\vskip2mm ~~\hskip10mm (a)\hskip60mm (b) \vskip2mm \caption{Total
cross section vs. $\sqrt{s}$ graphs for $Z_L$ associated pair
production of charged scalars at $f=5TeV$ when parameters
(a)$s/s':0.8/0.7$ and (b)$s/s':0.6/0.4$.} \label{PCM12x5}
\end{center}
\end{figure}


For the associated pair production of doubly charged scalars within
$Z_L$ the differential cross sections with respect to $E_Z$ are
plotted in figure \ref{zppC1a}(b) for different fixed values of
mixing angles at $\sqrt{s}=3TeV$, at symmetry breaking scale
$f=1TeV$. The dependance of total cross section on the center of
mass energy of this production process is plotted in figure
\ref{PCM12}, and the numerical values of total cross sections are
presented in table \ref{cseccharged} for parameters of interest. The
differential cross section of the production process reaches its
maximum when model parameters $s/s'=0.5/0.1$ about $0.1\fbg$ for
$E_Z\sim100GeV$, resulting a total cross section of $84 fb$. This
will give about $8000$ events per year for high luminosities such as
$100fb^{-1}$. But this remarkable event numbers are out of reach,
since $s/s'=0.5/0.1$ is out favored by electroweak presicion data
for $f=1TeV$. For the electroweak allowed parameters
$s/s'=0.8/0.6,0.8/0.7,0.95/0.6$ at $f=1TeV$ at $\sqrt{s}=3TeV$, the
differential cross section gets low values at the order of
$10^{-4}\fbg$. The resulting cross sections are calculated by
integrating over $E_Z$, and found about $0.4\sim0.8\times fb$ (table
\ref{cseccharged}) resulting $40\sim80$ events per year for
integrated luminosity of $100fb^{-1}$. For $\sqrt{s}<2TeV$, this
production channel is not reachable. In the {\lsth} model,
$\phi^{++}$ has decays to charged vectors $W^+_L W^+_L$ and also to
leptons $l^+_i l^+_j$ proportional to squares of the values of the
yukawa couplings; $|Y^2|$ for same families and $|Y'^2|$ for
different lepton families when lepton violating modes are
considered. So this channel provides final signals for doubly
charged scalar discovery and lepton flavor violation. The final
states of the doubly charged scalar pairs dominantly contain
leptonic modes $l_i l_i l^+_i l^+_j $, semi leptonic modes $l_i l_i
W^+_L W^+_L$ and to standard model charged vector pair $W^+_L W^+_L
W^-_LW^-_L$ depending on $Y$ and $Y'$, while $Z_L$ dominantly decays
to jets carrying the energy at the order of masses of the scalar
pair. For $f=1TeV$ the leptonic branching ratio of doubly charged
scalars can reach values close to $1$ for $Y\rightarrow 1$,
independent from $Y'$. If the value of the yukawa coupling $Y$ is
high enough $(Y\sim 1)$, the number of final state lepton flavor
violating signals such as; $Z_L l_il_i l^+_j l^+_j$, can reach up to
$50$ events per year for luminosities of $100fb^{-1}$, which can be
directly detectable free from backgrounds.

Finally, we have also analyzed the behavior of the production
processes for higher values of $f$. The dependence total cross
section on $\sqrt{s}$ when $s/s'=0.8/0.7$ for $f=2.5TeV, 3.5TeV,
 5TeV$ are plotted in figures \ref{PCM12x2}(a), \ref{PCM12x2}(b)
and \ref{PCM12x5}(a), respectively. It is seen that the maximum
value of total cross sections, production rates and final lepton
flavor violating signals for these $f$ values for both processes
remain same, but the required energy is shifted to $\sqrt{s}=3.8TeV
(5.5TeV) \{7.5TeV\}$ for $f=2.5TeV (3.5TeV) \{5TeV\}$, due to the
increase in heavy scalar masses. For $f=5TeV$, we have also analyzed
the case $s/s'=0.6/0.4$, since the parameters are less constrained.
The dependence of total cross sections for associated production of
both single and doubly charged pairs are plotted in figure
\ref{PCM12x5}(b). It is seen that the total cross sections are
increased by order one. For the single charged pair, the production
cross section is calculated as $0.9 fb$ for $\sqrt{s}\gtrsim 10TeV$,
resulting $90$ production events for luminosities $100 fb^{-1}$. For
the double charged scalar pair, the total cross section is $5\sim 9
fb $ for $\sqrt{s}\gtrsim 10TeV$, giving $500\sim900$ productions at
luminosities $100fb^{-1}$. In this case the final number of four
lepton signals ($l_i l_i l^+_i l^+_j $) will be around $600$ for
higher values of yukawa coupling ($Y\sim 1$). If the values of
yukawa coupling is smaller, in the region $0.1\leq Y\leq 0.3$, the
double charged pair will decay into semi leptonic modes, such as
$W^+_L W^+_L l_i l_i$, resulting $200\sim400$ signals accessible for
luminosities of $100fb^{-1}$.

In conclusion, we found that at an $e^{+}e^{-}$ collider of
$\sqrt{s}\geq 2TeV$ with a luminosity of $100fb^{-1}$, the $Z_L$
associated production of charged scalars will be in the reach, being
the single charged pair is quite challenging due to low production
rates, and the production of doubly charged scalar pair more
promising for the electroweak allowed parameters at $f=1TeV$. The
final states will contain lepton flavor violating signals if the
value of yukawa coupling $Y$ is close to unity. For larger values of
$f$ the mixing angles $s/s'$ are less constrained, e.g. for $f=5TeV$
and $s/s'=0.6/0.4$, the production rates increases allowing
remarkable final lepton number violating events for $0.1\leq Y \leq
1$, but for these set of parameters the center of mass energy of the
colliders should be increased.
\begin{acknowledgments}
The authors would like to thank to T. M. Aliev for useful
discussions and his support during this work. A. \c{C}. also thanks
to A. \"Ozpineci for his support and guidance.
\end{acknowledgments}


\clearpage

\end{document}